\documentclass[twocolumn]{article}

\usepackage[utf8]{inputenc}
\usepackage{authblk}
\usepackage{graphicx}
\usepackage{amsmath}
\usepackage{amsfonts}
\usepackage{amssymb}
\usepackage{hyperref}
\usepackage{multirow} 

\begin{document}

\title{Sliding Dumbbell Method to search for the CME in heavy ion collisions}


\author[1]{Madan M. Aggarwal}
\author[1]{Anjali Attri}
\author[1,2]{Sonia Parmar}
\author[1,3]{Anjali Sharma}
\author[1]{Jagbir Singh}

\affil[1]{Department of Physics, Panjab University, Chandigarh, India 160014}
\affil[2]{Present address: 3732 AR, Utrecht, Netherlands}
\affil[3]{Present address: Department of Physical Sciences, BOSE Institute, Kolkata, India 700091}
\affil[ ]{Corresponding author: aggarwal@pu.ac.in}

\twocolumn[{
    
\maketitle

\begin{abstract}
  This study explores the Chiral Magnetic Effect (CME) in ultra-relativistic heavy-ion collisions. The CME, observed as back-to-back charge separation along the magnetic field axis, is investigated using the newly developed Sliding Dumbbell Method (SDM) applied to Au+Au events at a center-of-mass energy $\sqrt{s}_{\mathrm{NN}}$ = 200 GeV generated by the AMPT model with string melting configuration. The CME-like signal is externally injected in events by flipping charges of pairs of the particles perpendicular to the reaction plane. The study reports a significant enhancement of the CME-sensitive 3-particle $\gamma$ correlator in events with high back-to-back charge separation, in a given collision centrality. Additionally, a linear relationship is observed between the $\sqrt{|\gamma|}$ correlator for same-sign charge pairs and positive charge asymmetry ($\langle A{^+}\rangle$) across the dumbbell in CME-enriched sub-samples. Furthermore, the fraction of CME in $\Delta\gamma$ (difference between opposite and same sign $\gamma$ correlators) is presented across different collision centralities having different percentages of externally injected CME-like signal. Overall, the research aims to understand and detect the Chiral Magnetic Effect through innovative experimental method and detailed analysis of event structure. 
\end{abstract}
\\



\textbf{pacs:} {25.75.-q; 12.38.Mh}\\ 

}]



\section{Introduction}\label{sec:intro}
In the realm of strong interactions, parity conservation holds true under conditions of vanishing temperature and isospin density~\cite{Vaf84}. However, there is a proposition suggesting that parity may experience local violations within microscopic domains of quantum chromodynamics (QCD) at finite temperatures. In the vicinity of the Quark Gluon Plasma (QGP) in ultra-relativistic heavy-ion collisions, occurring within a hot and dense medium, metastable domains characterized by gluon fields with non-zero topological charge could emerge~\cite{Lee73,Lee74,Kha98}. These domains interact with chiral quarks, inducing an asymmetry between left- and right-handed quarks. This chiral imbalance, coupled with the presence of a strong magnetic field (generated by energetic spectator protons in non-central heavy-ion collisions), results in charge separation along the magnetic field axis, a phenomenon known as the Chiral Magnetic Effect (CME)~\cite{Kha06,Kha07,Kha08,Fuk08}. Charged hadrons originating from quark hadronization may exhibit experimentally detectable charge separation perpendicular to the reaction plane. Currently, the CME is the subject of intense theoretical and experimental scrutiny in relativistic heavy-ion collisions, notably at the Relativistic Heavy Ion Collider (RHIC) at BNL and the Large Hadron Collider (LHC) at CERN~\cite{Bzd13,Kha16,Koc17,Sch19,Zha19}.\par
The CME-like charge separation can be investigated via the first P-odd sine term in a Fourier decomposition of the azimuthal distribution of charged particles~\cite{Vol04,Vol90}:
\begin{equation}
  \begin{split}
    \frac {dN^{a}}{d\phi} \propto 1+2v_{1,a}cos(\Delta\phi^{*})+2v_{2,a}cos(2\Delta\phi^{*})...\\
    +2a_{1,a}sin(\Delta\phi^{*})+...
    \label{eq:eq1}
  \end{split}
\end{equation}
 where $\Delta\phi^{*}$ = $\phi -\Psi_{RP}$, $\phi$ is the azimuthal angle of the particle and $\Psi_{RP}$ is the reaction plane angle; $v_{1}$ and $v_{2}$ are coefficients referred to as directed and elliptic flows, respectively; $a_{1}$ represents the parity violating effect, and $a$ indicates type of particle. These coefficients depend on the transverse momenta and rapidities of particles. The coefficient, $a_{1}$, vanishes when averaged over many events, because if spontaneous parity violation occurs, sign of $a_{1}$ will vary from event to event, leading to $\langle a_{1} \rangle$ = 0. However, one can observe the effect via the CME-sensitive $\gamma$ correlator~\cite{Vol04} defined as:
\begin{equation}
  \begin{split}
    \gamma& =\langle cos (\phi_{a}+\phi_{b}-2\Psi_{RP}) \rangle \\
    & =\langle cos (\Delta\phi_{a}^{*})cos(\Delta\phi_{b}^{*}) \rangle - \langle sin(\Delta\phi_{a}^{*}) sin(\Delta\phi_{b}^{*}) \rangle \\
    & =v_{1,a}v_{1,b} - a_{1,a}a_{1,b}
    \label{eq:eq2}
  \end{split}
\end{equation}
       where $\phi_{a}$ and $\phi_{b}$ represent the azimuthal angles of the particles ``a'' and ``b'', respectively. Here, average is taken over particles in an event and over the events in a given sample. The $\gamma$ correlator represents the difference between the in-plane ($\langle cos(\Delta\phi_{a}^{*}) cos( \Delta\phi_{b}^{*})\rangle$) and the out-of-plane ($\langle sin(\Delta\phi_{a}^{*}) sin(\Delta\phi_{b}^{*}) \rangle$) correlations. The out-of-plane term is sensitive to the CME-like charge separation (i.e., $a_{1,a}a_{1,b}$) whereas the in-plane cosine term serves as a reference since both are equally sensitive to backgrounds unrelated to the reaction plane~\cite{Bor02}. It is possible to measure the parity violation effects (i.e., $a_{1,a}a_{1,b}$) using Eq.(\ref{eq:eq2}) provided directed flow $v_{1}$ = 0. In order to compute in-plane and out-of-plane correlations, we also calculated the 2-particle $\delta$ correlator given as:
  \begin{equation}
    \begin{split}
      \delta& = \langle cos(\phi_{a} - \phi_{b}) \rangle \\
      & = \langle cos(\Delta\phi_{a}^{*}) cos( \Delta\phi_{b}^{*}) \rangle + \langle sin(\Delta\phi_{a}^{*}) sin(\Delta\phi_{b}^{*}) \rangle
      \label{eq:eq3}
    \end{split}
  \end{equation}
        In heavy-ion collisions, the reaction plane angle is not known, so one usually estimates the second order event plane ($\Psi_{2}$) from particles' azimuthal angles. Therefore, the $\gamma$ correlator is determined using $\Psi_{2}$ instead of $\Psi_{RP}$ and is corrected by applying the event plane resolution. However, Voloshin~\cite{Vol04} proposed the following equation if the event plane is determinated using only one particle,
  \begin{equation}
    \langle cos(\phi_{a} + \phi_{b} - 2\phi_{c}) \rangle = (v_{1,a}v_{1,b} - a_{1,a}a_{1,b}) v_{2,c}
    \label{eq:eq4}
  \end{equation}
  here particle ``c'' is used to determine the event plane and $v_{2,c}$ is its elliptic flow. The STAR collaboration~\cite{Abe09,Abe10} at RHIC demonstrated that, within errors, it is approximately equal to the $\gamma$ correlator defined in the Eq.(\ref{eq:eq2}) as:
  \begin{equation}
    \gamma = \langle cos(\phi_{a} + \phi_{b} - 2\Psi_{RP}) \rangle \approx \frac {\langle cos(\phi_{a} +\phi_{b} -2\phi_{c}) \rangle} {v_{2,c}}
    \label{eq:eq5}
  \end{equation}
The $\gamma$ and $\delta$ correlators are determined for same and opposite sign charge pairs averaged over particles in an event followed by averaging over events. The difference between $\gamma$ correlators for opposite ($\gamma_{OS}$) and same sign ($\gamma_{SS}$) charge pairs is also estimated as:
$\Delta\gamma$ = $\gamma_{OS}$ - $\gamma_{SS}$.
In- and out-of-plane correlations are determined using Eqs.(\ref{eq:eq2}) and~(\ref{eq:eq3}) for the different charge combinations.\par
In the case of the Chiral Magnetic Effect (CME), several expectations arise~\cite{Bzd13}:
\begin{itemize}
\item $ |\gamma_{OS}| \approx |\gamma_{SS}|$: This equality reflects the charge separation effect expected from the CME phenomenon, where both OS and SS charge pairs contribute significantly to the overall correlation signal.
\item $\Delta\gamma > 0$: A positive value of $\Delta\gamma$ is a characteristic signature of the CME.
\item Out-of-plane correlations larger than in-plane correlations: This expectation arises from the fact that the CME induces a charge separation predominantly perpendicular to the reaction plane. Consequently, correlations stemming from this out-of-plane charge separation are expected to be more pronounced than correlations originating from in-plane charge dynamics.
\end{itemize}\par
The STAR~\cite{Abe09,Abe10,Ada13,Ada14} and the ALICE~\cite{Abe13} reported charge separation signal using $\gamma$ correlators. The CMS~\cite{Sir18} has put an upper limit on the CME contribution to $\Delta\gamma$ of 7\% in Pb+Pb collisions at $\sqrt{s}_{\mathrm{NN}}$ = 2.76 TeV at 95\% confidence level. The ALICE~\cite{Ach18} also reported an upper limit of 26-33\% at 95\% confidence level on the contribution of the CME signal in Pb+Pb collisions at $\sqrt{s}_{\mathrm{NN}}$ = 2.76 TeV using the Event Shape Engineering (ESE) technique. Recently, in another measurement of charge separation, the ALICE~\cite{Ach20} has put an upper limit, on the fraction of the CME signal ($f_{CME}$), of 15-18\% (20-24\%) at 95\% (99.7\%) confidence level for the 0-40\% collision centralities using the data driven background. Recently, the STAR collaboration reported some indication of finite CME signal in Au+Au collisions at $\sqrt{s}_{\mathrm{NN}}$ = 200 GeV using the participant and spectator planes~\cite{Spec}. However, the STAR collaboration did not observe the expected enhancement due to the CME signal in the Ru+Ru collisions over the Zr+Zr collisions in the non-central collisions at $\sqrt{s}_{\mathrm{NN}}$ = 200 GeV~\cite{Iso}.\par
In this article, we present the Sliding Dumbbell Method (SDM), designed to search minutely for back-to-back charge separation on an event-by-event basis in heavy-ion collisions. It is noted that the environment in each heavy-ion collision may not be conducive to give rise to the CME. In view of this, it is extremely crucial to search for its signal in each and every event in order to isolate potential events exhibiting back-to-back charge separation rather than looking for the signal averaging over events in a given collision centrality. An attempt is made to find the sample of events enriched with the CME-like signal in a given collision centrality employing the new method. The back-to-back charge separation can also happen due to statistical fluctuations giving rise to CME-like effect. Such type of back-to-back charge separation contributing to CME-like effect is obtained by shuffling the charges of particles in an event. The SDM is tested on AMPT generated events. Also, we report, for the first time, the linear dependence of $\sqrt{|\gamma|}$ correlator for the same sign charge pairs on the positive charge asymmetry ($\langle A{^+}\rangle$) across the dumbbell for the CME-enriched sub-samples.
\begin{figure}[h]
  \centering
  \includegraphics[width=.9\columnwidth]{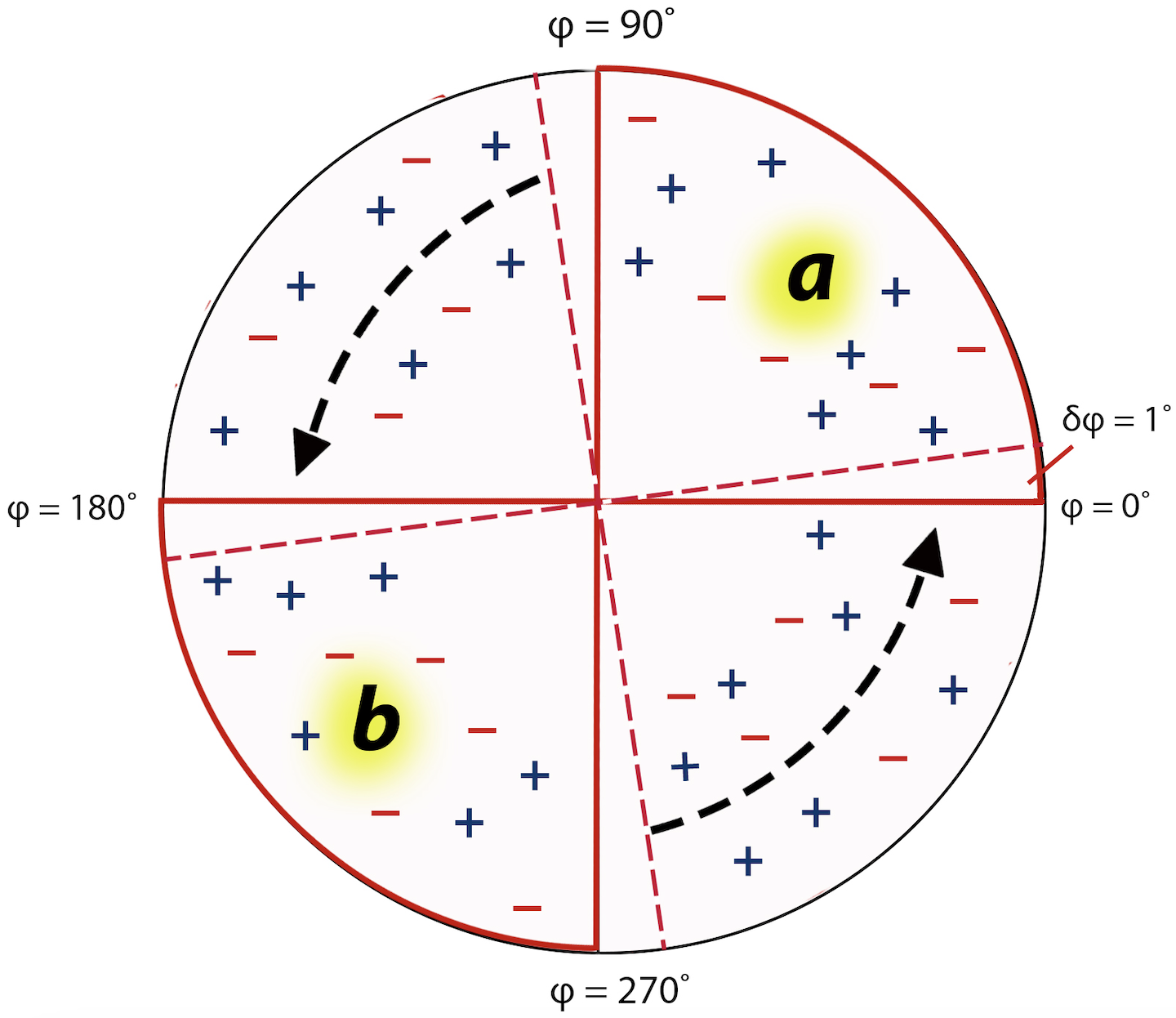}
  \caption{Pictorial depiction of transverse (azimuthal) plane with hits of positive (+) and negative (-) charge particles in an event. The red coloured dumbbell of size $90^{\circ}$ with ``a'' and ``b'' sides is also displayed. The black arrows indicate the anti clockwise sliding of the dumbbell and the dashed dumbbell slid by $\delta\phi$ = $1^{\circ}$ is also shown.}
  \label{fig1}
\end{figure}
\section{Sliding Dumbbell Method}
Fig.\ref{fig1} displays hits of positive and negative charge particles in an event represented by ``+'' and ``-'' symbols, respectively, on the transverse (azimuthal) plane i.e., plane perpendicular to the beam direction. The dumbbell is also shown using a solid red colour covering the azimuthal angles $0^{\circ}$-$90^{\circ}$ on one side and $180^{\circ}$-$270^{\circ}$ on the other side denoted as ``a'' and ``b'' sides, respectively. Below we define some terms associated with the dumbbell:
\begin{itemize}
\item The sum of the positive and the negative charge fractions across the dumbbell is defined as: 
\begin{equation}
Db_{+-} = \frac{n_{+}^{a}}{(n_{+}^{a} +n_{-}^{a})} +  \frac{n_{-}^{b}}{(n_{+}^{b} +n_{-}^{b})}
\label{eq:eq6}
\end{equation}
where $n_{+}^{a}$ ($n_{-}^{a}$) and $n_{+}^{b}$ ($n_{-}^{b}$) are the number of positive (negative) charge particles, in $90^{\circ}$ azimuth, on ``a'' and ``b'' sides of the dumbbell, respectively. The $Db_{+-}$=2 corresponds to 100\% back-to-back charge separation whereas $Db_{+-}$=1 means no back-to-back charge separation.
\item The charge excess asymmetry, $Db_{+-}^{asy}$, across the dumbbell is defined as:
\begin{equation}
      Db_{+-}^{asy}  = \frac{(N_{+}^{ex} -N_{-}^{ex})}{(N_{+}^{ex} + N_{-}^{ex})}
 \label{eq:eq7}
\end{equation}
where $N_{+}^{ex}$ ($=n_{+}^{a}-n_{-}^{a}$) is the positive particles' excess on the ``a'' side of the dumbbell and $N_{-}^{ex}$ ($=n_{-}^{b}-n_{+}^{b}$) is the negative particles' excess on the ``b'' side of the dumbbell.
 \item The positive charge asymmetry across the dumbbell in an event is given as:
\begin{equation}
  A^{+} = \frac{|n_{+}^{a}-n_{+}^{b}|}{N^{+}}
 \label{eq:eq8}
\end{equation}
\end{itemize}
The A$^{+}$ is related to the coefficient, $a_{1}$~\cite{Kha05}, describing the parity violating effect in Eq.(\ref{eq:eq1}). $N^{+}$ ($N^{-}$) represents the number of positive (negative) charge particles in an event.\par
In the Sliding Dumbbell Method, whole transverse plane (Fig.\ref{fig1}) is scanned minutely to search for the maximum value of the $Db_{+-}$ by sliding the dumbbell in steps of $\delta\phi=1^{\circ}$ as shown in the figure by black arrows while calculating $Db_{+-}$ and $Db_{+-}^{asy}$ for each slid setup. In other words, we obtain 360 values of $Db_{+-}$ and $Db_{+-}^{asy}$ using Eqs.(\ref{eq:eq6}) and (\ref{eq:eq7}), respectively, for different azimuthal coverages on ``a'' and ``b'' sides (i.e.,
$0^{\circ}$-$90^{\circ}$, $180^{\circ}$-$270^{\circ}$; $1^{\circ}$-$91^{\circ}$, $181^{\circ}$-$271^{\circ}$;.........; $359^{\circ}$-$89^{\circ}$, $179^{\circ}$-$269^{\circ}$). Out of these 360 values of $Db_{+-}$, we find maximum value of $Db_{+-}$ i.e., $Db_{+-}^{max}$, with the condition $|Db_{+-}^{asy}|<0.25$ in order to select the events similar to the CME-like events rather than one sided either positive or negative particles' excess across the dumbbell. The above procedure is repeated for each event to get $Db_{+-}^{max}$. The back-to-back fractional charge separation across the dumbbell, $f_{DbCS}$, can be written as:
  \begin{equation}
    f_{DbCS} = Db_{+-}^{max} -1
    \label{eq:eq9}
  \end{equation}
  Hereinafter, $f_{DbCS}$ is referred to as back-to-back charge separation. For each collision centrality, we obtain $f_{DbCS}$ distribution. In order to get the CME-like enriched sample, we sliced $f_{DbCS}$ distribution for each centrality into 10 percentile bins (i.e., 0-10\%, 10-20\%, ......, 90-100\%). The 0-10\% class corresponds to the most potential CME-like candidates with high back-to-back charge separation (i.e., higher $f_{DbCS}$ values) whereas 90-100\% class corresponds to low back-to-back charge separation (i.e., lower $f_{DbCS}$ values). We study the CME-sensitive $\gamma$ and $\delta$ correlators for each sliced $f_{DbCS}$ bin in each centrality to investigate their dependencies on charge separation across the dumbbell.
  
  \section{Background Estimation}
  To ascertain the background contribution in each sliced $f_{DbCS}$ bin, the charges of particles in each event for a specific centrality are randomly shuffled. This shuffling destroys charge-dependent correlations among charged particles while preserving the polar angle ($\theta$) and azimuthal angle ($\phi$) of each particle within an event, ensuring that flow is not affected. Subsequently, the $f_{DbCS}$ distribution is derived for the charge-shuffled events at a given collision centrality. The $\gamma$ value for the charge-shuffled events' sample is then calculated for a sliced $f_{DbCS}$ bin, representing the contribution due to statistical fluctuations, and termed as $\gamma_{ChS}$.\par

To restore the correlations among particles, which were destroyed during charge shuffling, the $\gamma$ correlator is calculated from the corresponding events in the original events' sample for the sliced $f_{DbCS}$ bin of charge-shuffled events. This is termed as $\gamma_{Corr}$. Since $\gamma_{Corr}$ is derived from the original events themselves, it encompasses all types of correlations, such as resonance decays, flowing clusters, and transverse momentum conservation.\par
The background contribution ($\gamma_{bkg}$) to $\gamma$ is estimated as sum of $\gamma_{ChS}$ and $\gamma_{Corr}$:
\begin{equation}
    \gamma_{bkg}  =  \gamma_{ChS}  + \gamma_{Corr}
    \label{eq:eq10}
\end{equation}

\section{Data Samples}
We generated 16.3 million Au+Au events at a center-of-mass energy $\sqrt{s}_{\mathrm{NN}}$ = 200 GeV using the AMPT Monte Carlo event generator with string melting configuration. During the generation of AMPT events, the reaction plane angle ($\Psi_{RP}$) was set to zero, and these events were categorized into various collision centralities (ranging from 0-10\% for the most central events to 90-100\% for peripheral events) based on the total number of participating nucleons. Within each event, we applied experimental track cuts~\cite{Abe09,Abe10}, selecting tracks with transverse momenta ($p_{t}$) ranging from 0.15 to 2.0 GeV/\textit{c} and pseudorapidity ($\eta$) values within the range of -1 to 1, with full azimuthal coverage, for subsequent analysis. We measured the $\gamma$ correlator in a rapidity region symmetric about the mid-rapidity, where $v_1$ = 0, enabling the detection of parity violation effects (e.g., $a_{1,a}a_{1,b}$) using  Equation~\ref{eq:eq4}.\par
\begin{table}[!htb]
  \begin{center}
    \begin{tabular}{|c|c|c|}
      \hline
      Sr. No.  & {Collision centrality} & {Percentage of} \\ [3pt]
               & centrality & CME-like \\ [3pt]
               &  & injected signal \\ [3pt]
      \hline
      1. & 10-20\%  & $\sim$0.65$\%$\\
      2. & 20-30\%  & $\sim$0.95$\%$ \\
      3. & 30-40\%  & $\sim$1.4$\%$ \\
      4. & 40-50\%  & $\sim$2.2$\%$ \\
      5. & 50-60\%  & $\sim$4.0$\%$ \\
      6. & 60-70\%  & $\sim$6.7$\%$ \\
      \hline
    \end{tabular}
  \end{center}
  \caption{The percentage of CME-like injected signal in Au+Au AMPT at $\sqrt{s_{NN}}$ = 200 GeV for various collision centralities.}
  \label{table2}
\end{table}
To introduce a CME-like signal, we randomly selected two negatively charged particles within the azimuth range of $45^{\circ}$ to $135^{\circ}$ and flipped them to two positively charged particles. Similarly, within the azimuth range of $225^{\circ}$ to $315^{\circ}$, two positively charged particles were flipped to two negatively charged particles. This process creates a back-to-back charge separation perpendicular to the reaction plane (with $\Psi_{RP}$=0), as expected in the case of the CME. The resulting sample is referred to as the CME sample. It is important to note that while the number of flipped particles in each event remains constant, the CME signal (defined as the ratio of flipped particles to the total number of particles in an event) varies from event to event based on the total number of particles in each event. Consequently, the injected CME signal ranges from approximately 7\% for collision centralities of 60-70\% to around 0.6\% for centralities of 10-20\% (see Table 1). Additionally, we generated charge-shuffled event samples for both the CME and AMPT samples. The subsequent analysis was conducted on the following samples for each centrality category:\par
  \begin{figure*}[!h]
    \centering{
    \includegraphics[width=.8\columnwidth]{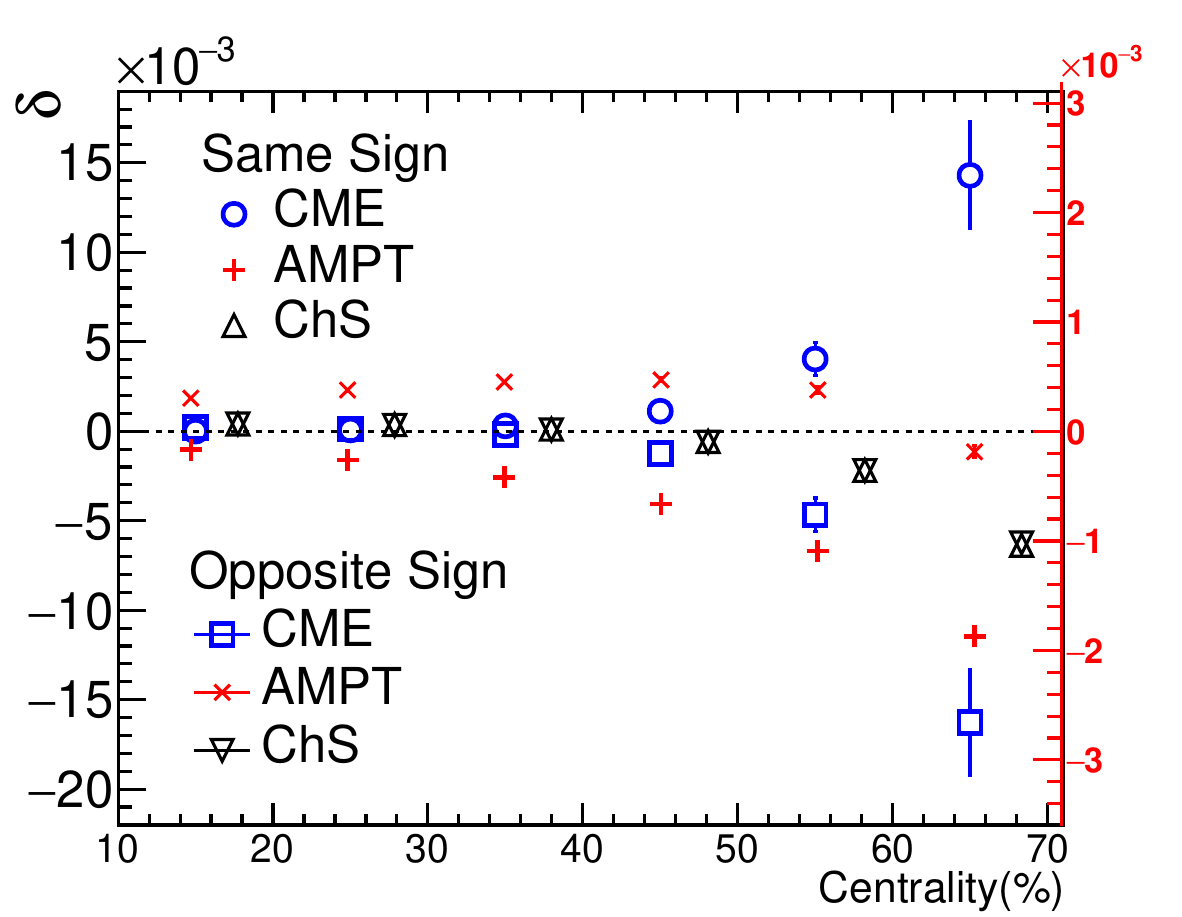}
    \includegraphics[width=.8\columnwidth]{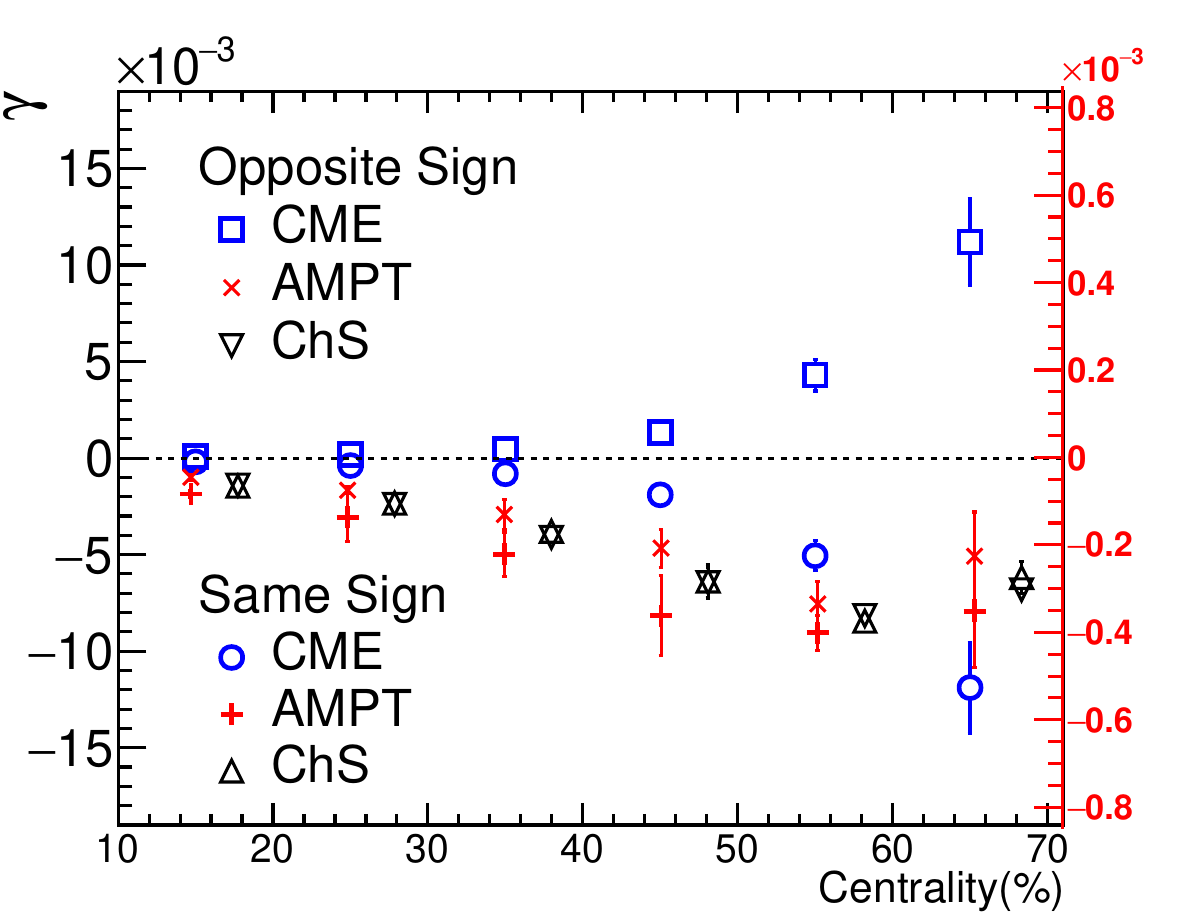}
    \includegraphics[width=.8\columnwidth]{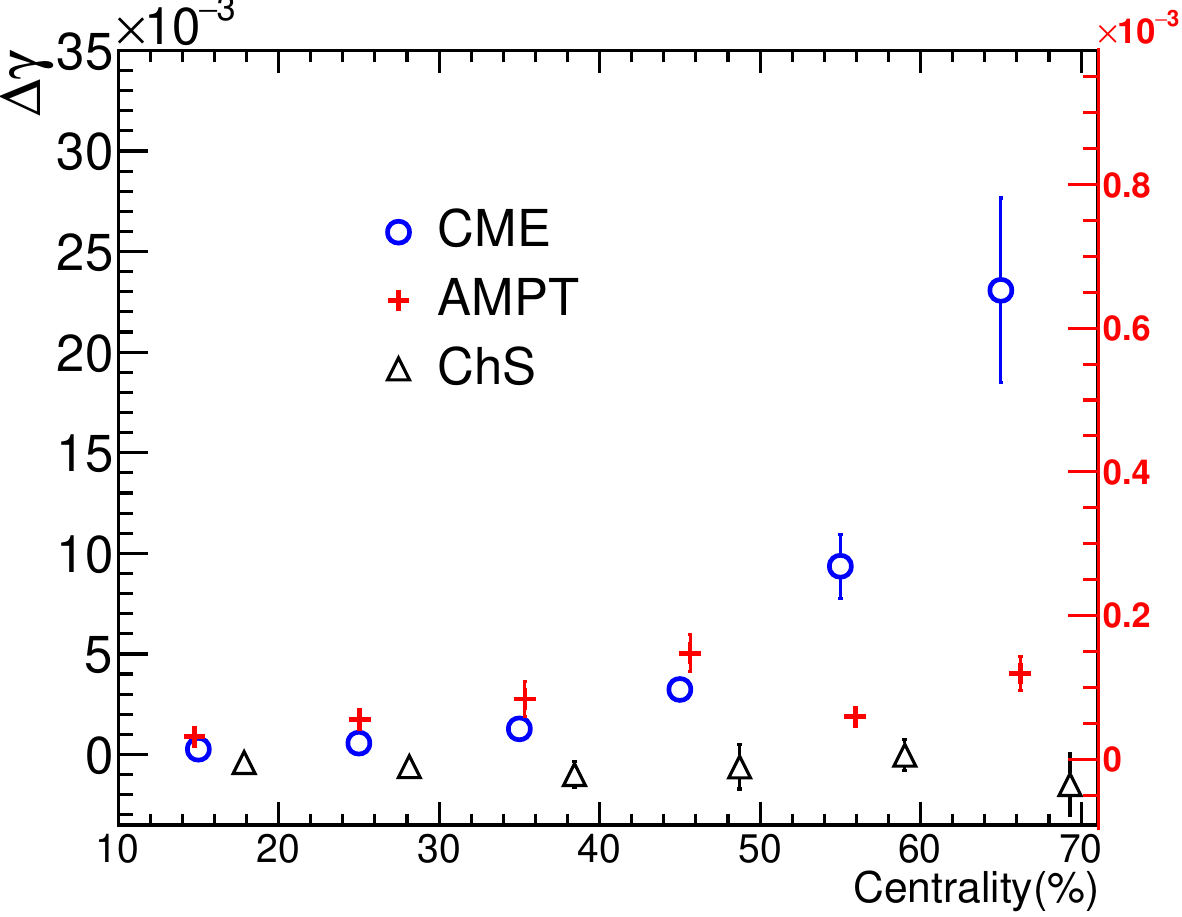}}
    \caption{Top Left: Centrality dependence of 2-particle $\delta$ correlator for same and opposite sign charge pairs for AMPT, CME and charge-shuffled (ChS) samples. Top Right: Similar plot for $\gamma$ correlator. Bottom: Centrality dependence of $\Delta\gamma$ for AMPT, CME and charge-shuffled (ChS) samples. For the AMPT and charge-shuffled (ChS) samples y-axes are given on the right hand sides.
    }
    \label{fig2}
  \end{figure*}
\begin{itemize}
\item AMPT (without CME-like signal injection)
\item CME (AMPT with CME-like signal injection) 
\item Charge-shuffled sample obtained from AMPT sample
\item Charge-shuffled sample obtained from CME sample
\end{itemize}
\section{Results and Discussion}
\begin{figure}[h!]
  \centering{
    \includegraphics[width=.75\columnwidth]{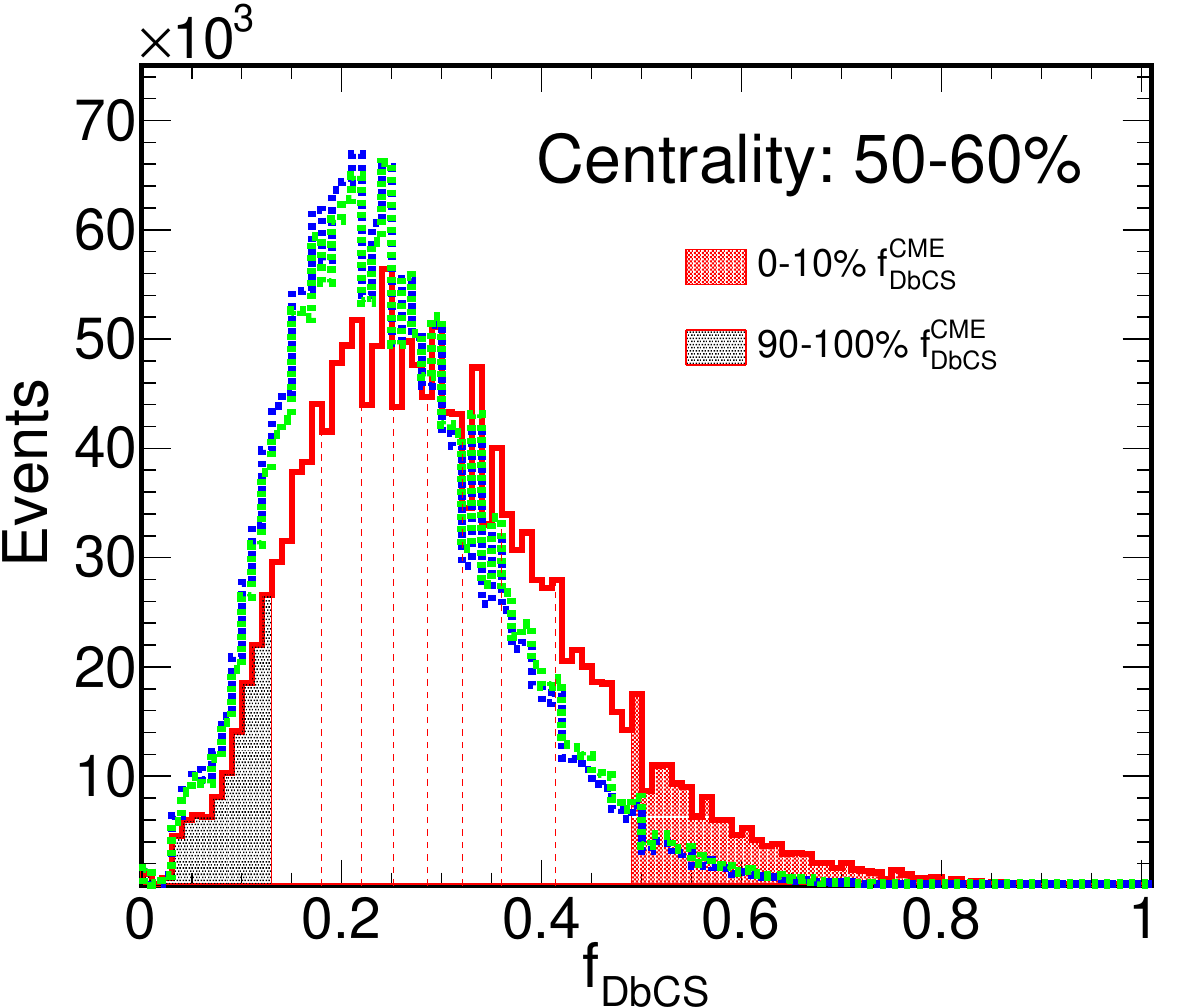}
    \includegraphics[width=.75\columnwidth]{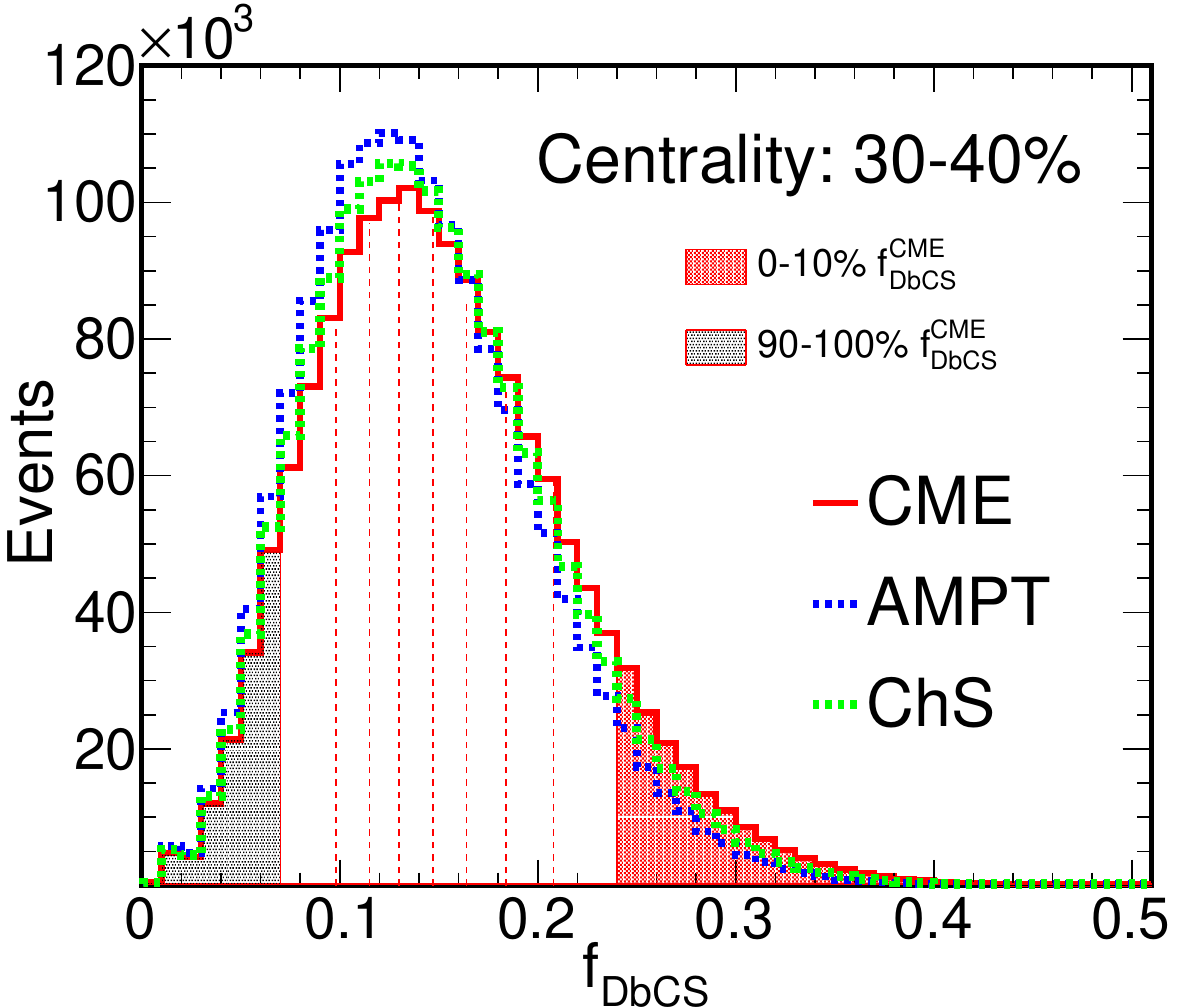}}
  \caption{Comparison of $f_{DbCS}$ distributions for 50-60\% (Top) and 30-40\% (Bottom) collision centralities as indicated in the figure for AMPT, charge-shuffled denoted as ChS and CME ($\sim$4\% injected signal for 50-60\% centrality and $\sim$1\% injected signal for 30-40\% centrality) samples. The vertical red lines divide $f_{DbCS}^{CME}$ distribution into 10 bins with roughly the same number of events.}
  \label{fig3}
\end{figure}
\begin{figure*}[h!]
  \centering{
    \includegraphics[width=.75\columnwidth]{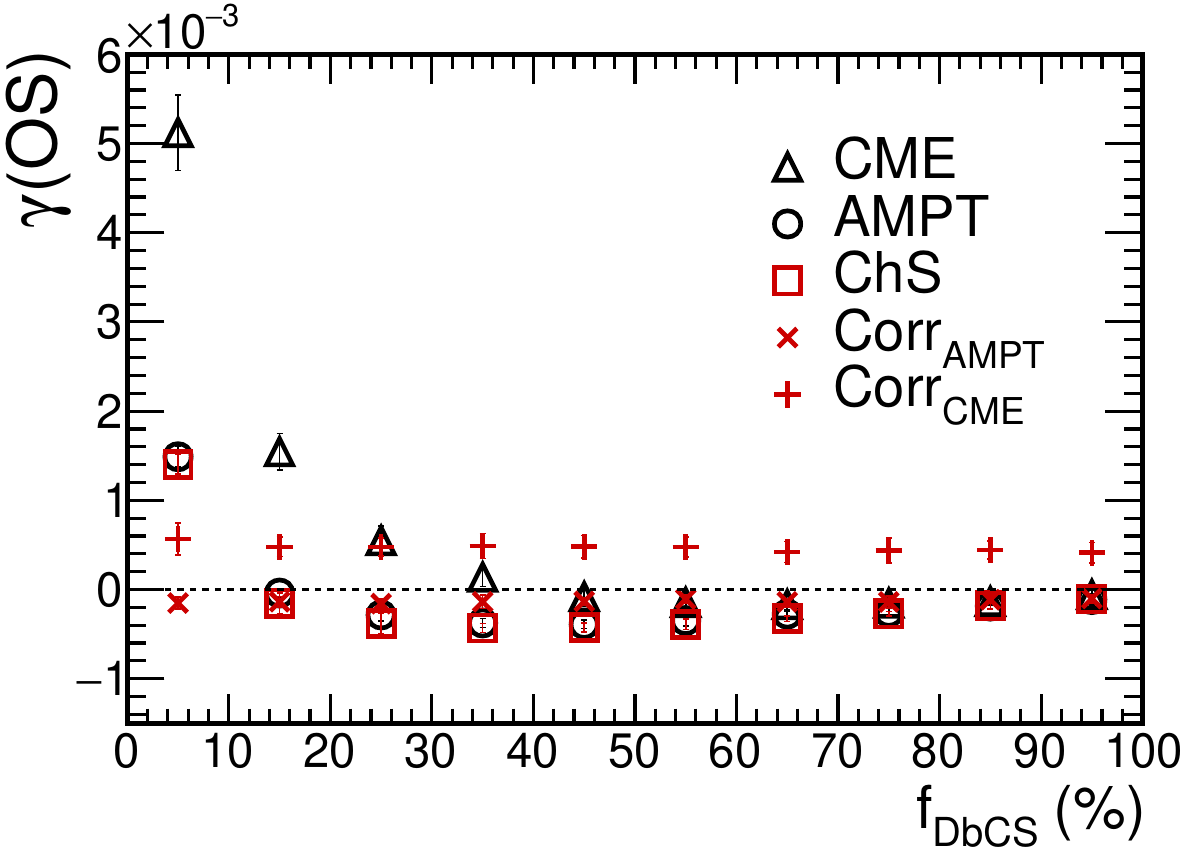}
    \includegraphics[width=.75\columnwidth]{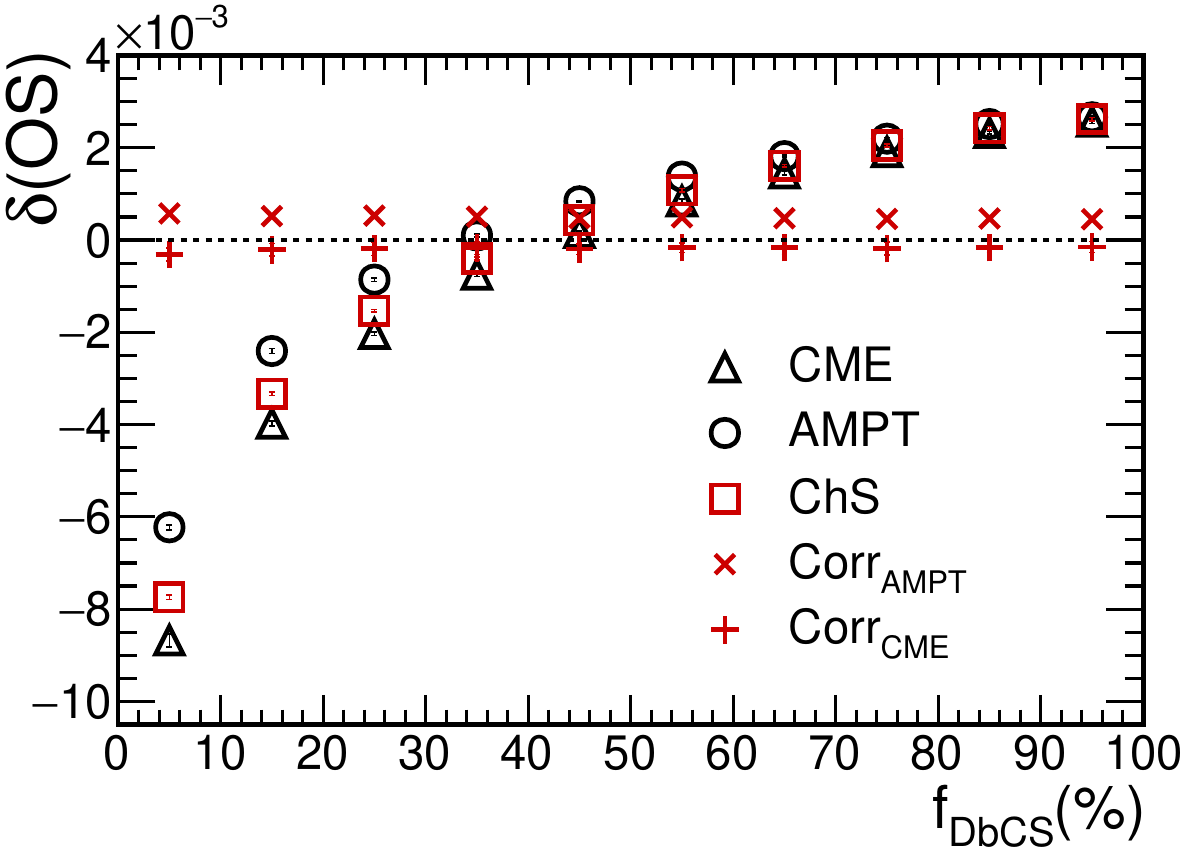}
    \includegraphics[width=.75\columnwidth]{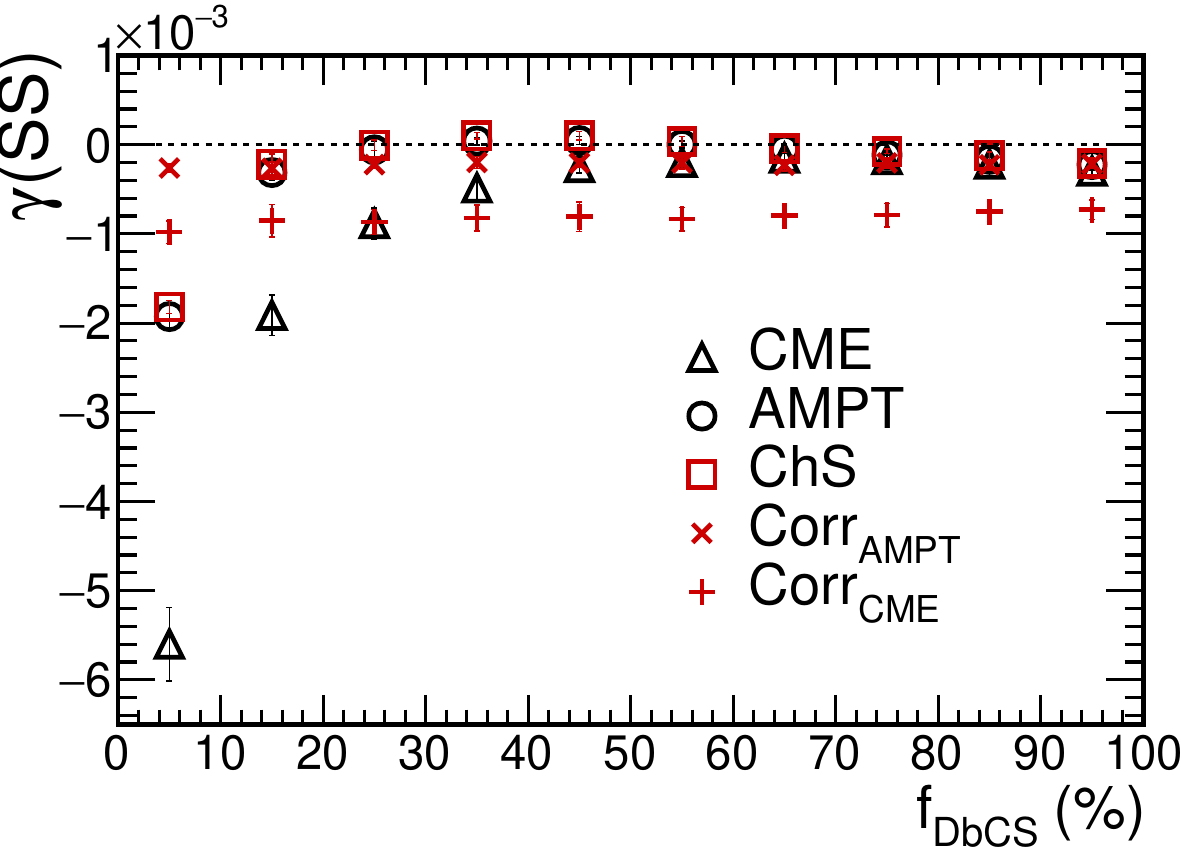}
    \includegraphics[width=.75\columnwidth]{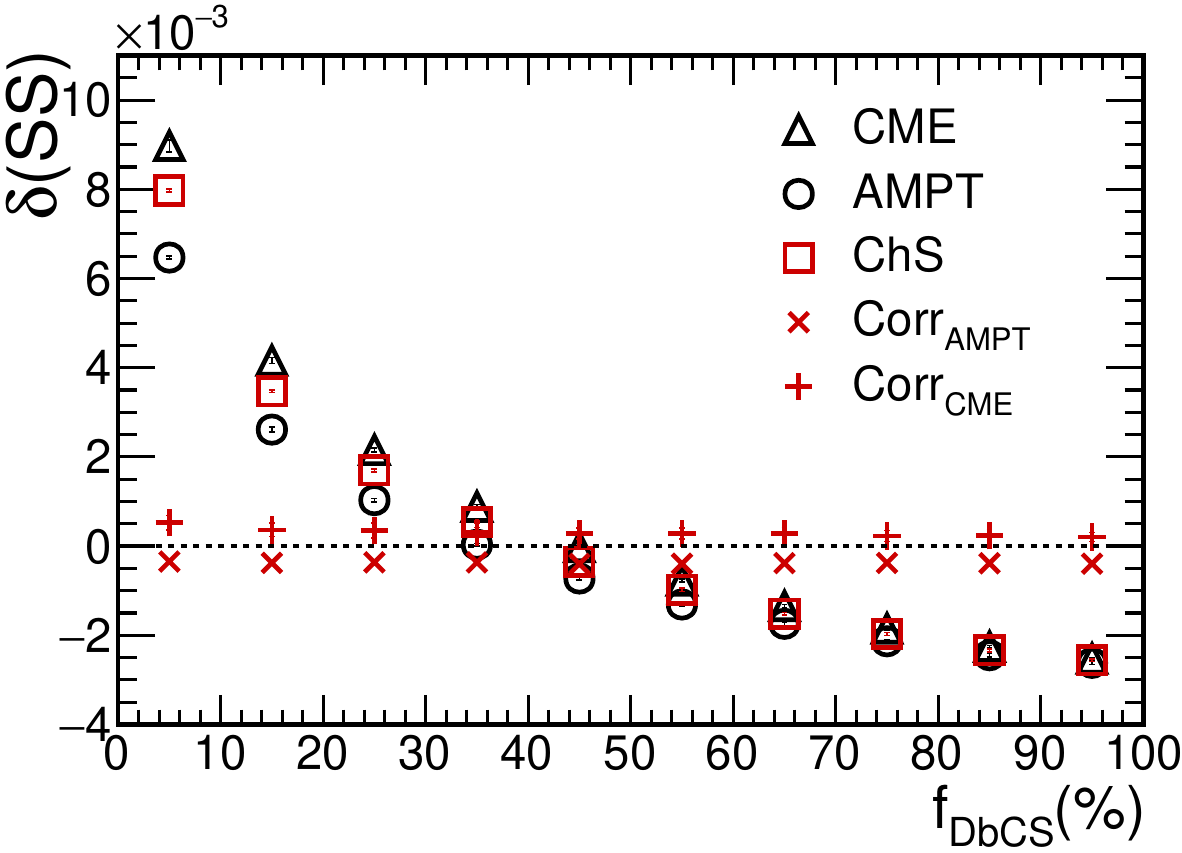}}
  \caption{Left: The dependence of $\gamma$ correlator on the charge separation ($f_{DbCS}$) for OS (top Left) and SS (bottom Left) charge pairs for 30-40\% collision centrality for CME ($\sim$1\% externally injected CME-like signal), AMPT, charge-shuffled (ChS) and respective correlated backgrounds represented as $Corr_{CME}$ and $Corr_{AMPT}$. Right: Similar plots for $\delta$.}
  \label{fig4}
\end{figure*}
The Q-cumulant technique~\cite{Bil11} is used for computing 2- and 3- particle correlators. Also, the $v_{2}$ is estimated from 2- and 4-particle Q-cumulants. The $v_{2}$ is taken as the average of $v_{2}$\{2\} and $v_{2}$\{4\} to calculate the 3-particle $\gamma$ correlator~\cite{Abe13}. The correlations for the positive-positive and the negative-negative charge pairs are found to agree with each other within the statistical uncertainties and are combined into one set of points, denoted as the same sign charge pairs' correlations. \par 
 The centrality dependence of the 2-particle $\delta$ correlator for different charge combinations for AMPT, CME and charge-shuffled samples are shown in Fig.\ref{fig2} (Top Left). As the correlation values are quite small for the AMPT and charge-shuffled samples, so for these samples, the correlation values of y-axis are given with magnified scales on the right hand side of each figure. Only the charge-shuffled points obtained from CME sample is shown in the figure for the sake of clarity as charge-shuffled points obtained from AMPT sample agree within errors to those of charge-shuffled points from CME sample. \par 
 It is observed that the opposite sign (OS) charge pairs' correlations are stronger and negative whereas the same sign (SS) charge pairs' correlations are positive. For both same and opposite sign charge pairs' correlations, the magnitude of correlations increases with decrease in the collision centrality. The centrality dependence of the 3-particle $\gamma$ correlator for OS and SS charge pairs is displayed in Fig.\ref{fig2} (Top Right) for AMPT (with no signal injection), CME, and charge-shuffled samples. Strong correlations are seen for the SS charge pairs whereas weaker correlations are observed for the OS charge pairs. The strength of correlations decreases with increasing collision centrality as the percentage of the externally injected signal decreases with increasing collision centrality. Further, it is worth noting that in case of the CME sample for the SS charge pairs, $\delta$ is positive and $\gamma$ is negative whereas for the OS charge pairs, $\delta$ is negative and $\gamma$ is positive. The correlations for the charge-shuffled are negative and have similar magnitude for both SS and OS charge pairs. Fig.\ref{fig2} (Bottom) exhibits $\Delta\gamma$ variation with collision centrality for different samples. For the AMPT events with no signal injection and charge-shuffled samples, $\Delta\gamma$ is small whereas for the CME sample, $\Delta\gamma$ has maximum value for 60-70\% collision centrality and it decreases with increasing collision centrality i.e., decreasing externally injected CME signal. \par 
\begin{figure*}[h!]
  \centering{
  \includegraphics[width=.75\columnwidth]{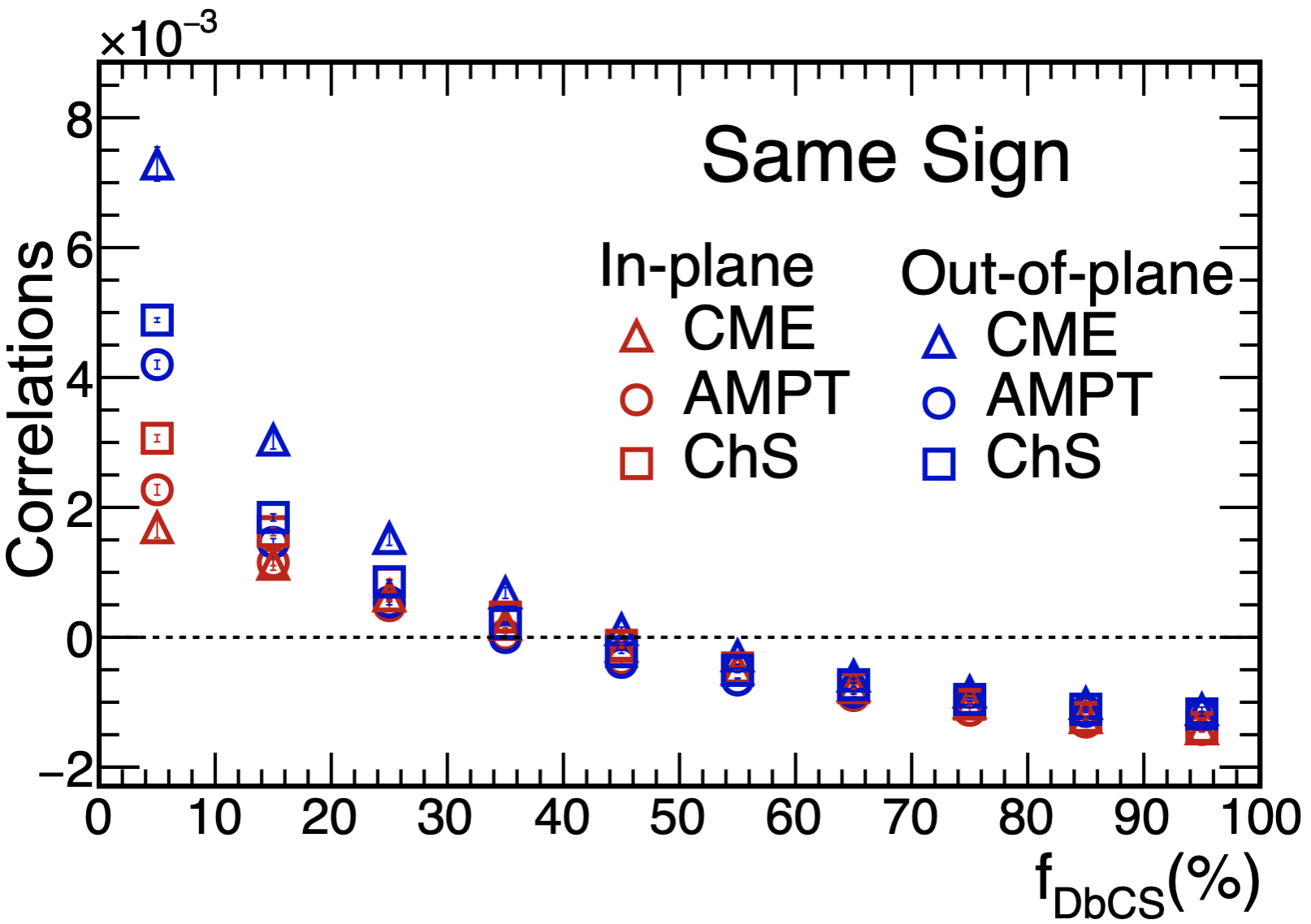}
  \includegraphics[width=.75\columnwidth]{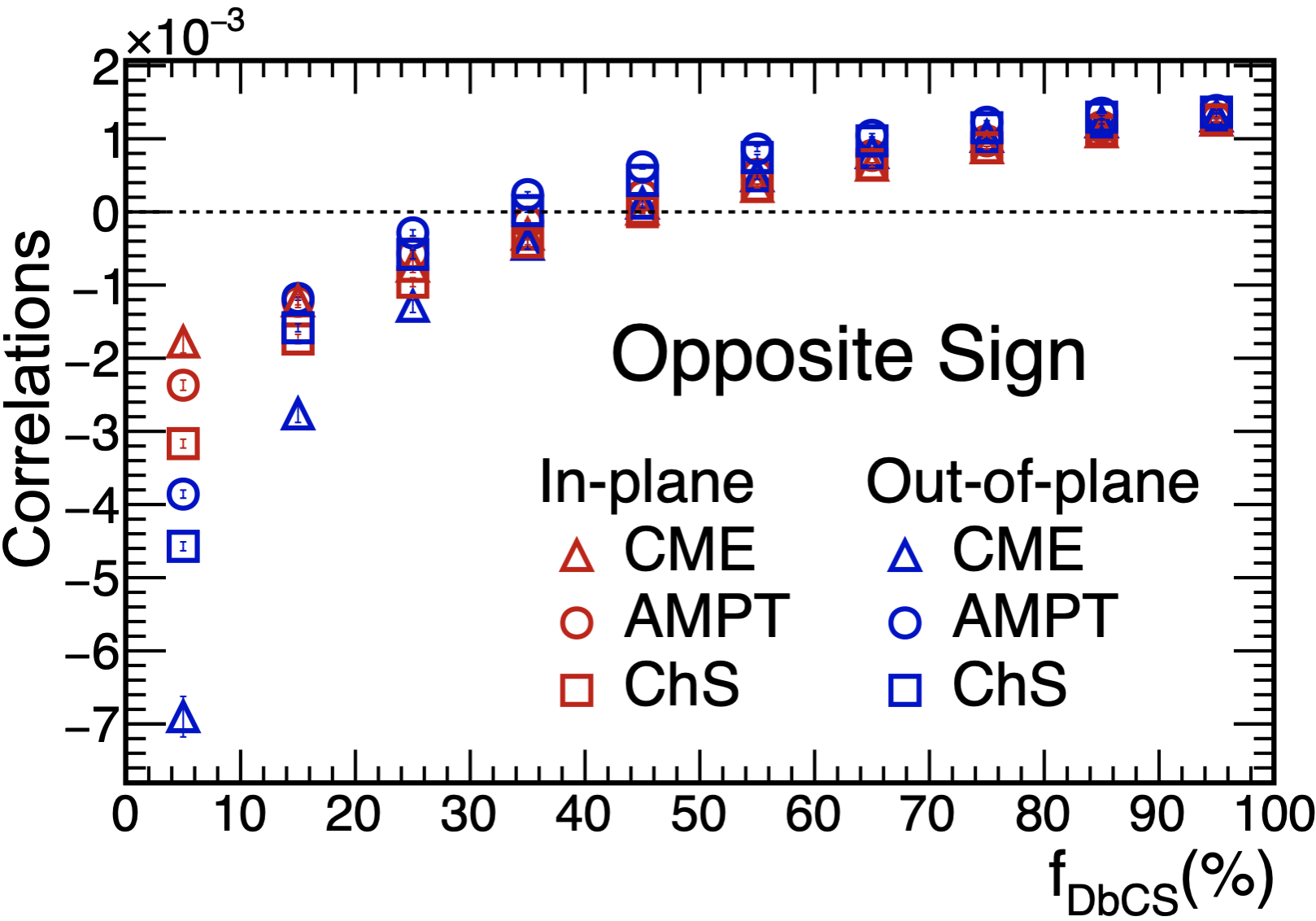}}
  \caption{Decomposition of the correlators into in-plane (red symbols) and out-of-plane (blue symbols) correlations for the same (Left) and the opposite (Right) sign charge pairs.}
\label{fig5}
\end{figure*}
\begin{figure}[!h]
  \centering{
  \includegraphics[width=.8\columnwidth]{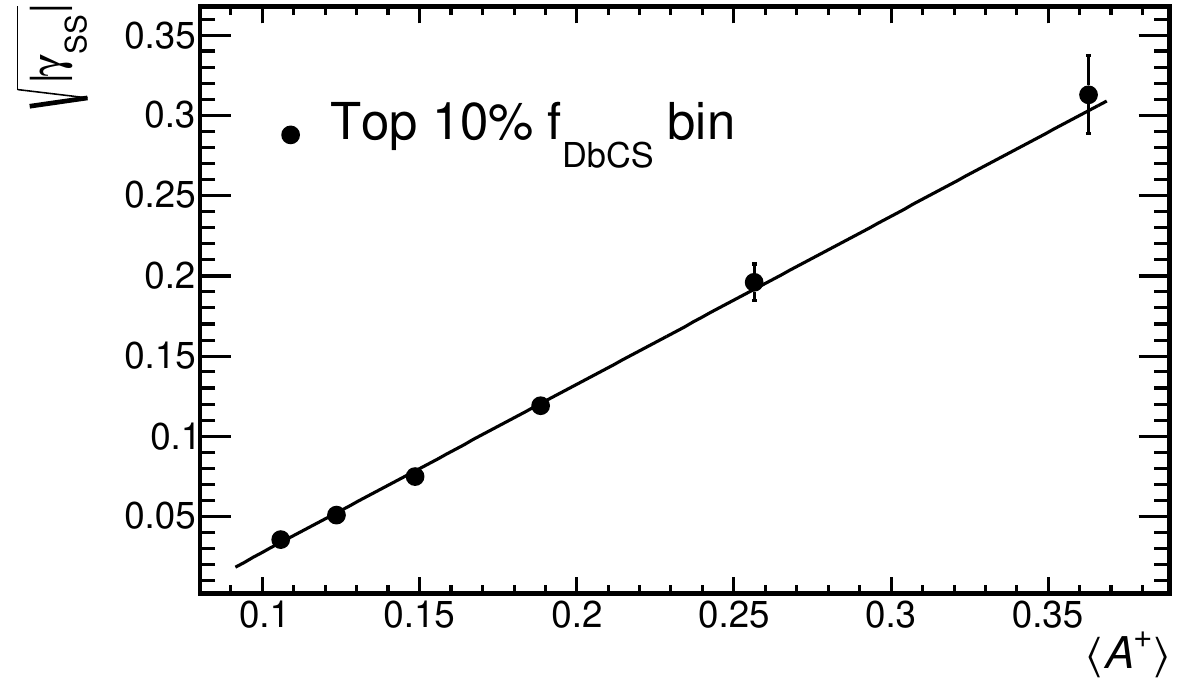}}
  \caption{The dependence of $\sqrt{|\gamma_{SS}|}$ on positive charge asymmetry ($\langle A^{+} \rangle$) across the dumbbell for the top 10\% $f_{DbCS}$ for the CME-enriched samples from each centrality.}
  \label{fig6}
\end{figure}
Now using the Sliding Dumbbell Method, we obtained the fractional charge separation ($f_{DbCS}$) across the dumbbell (Eq.(\ref{eq:eq9})) for each event. The $f_{DbCS}$ distributions for different samples i.e., AMPT, charge-shuffled and CME are compared in Fig.\ref{fig3} for the 50-60\% ($\sim$4\% CME signal) and 30-40\% ($\sim$1\% CME signal) collision centralities. The $f_{DbCS}$ distributions shift towards higher $f_{DbCS}$ values with higher percentage of externally injected CME-like signal for a given centrality. The $f_{DbCS}$ distributions move towards lower values of $f_{DbCS}$ with increasing collision centrality. It is noticed that $f_{DbCS}$ distributions with externally injected CME-like signal extend to higher $f_{DbCS}$ values than those of AMPT and charge-shuffled samples. Further, these $f_{DbCS}$ distributions are sliced into 10 percentile bins with 0-10\% bin containing top $f_{DbCS}$ values extending upto 1 and 90-100\% with lower $f_{DbCS}$ values approaching zero.\par
In Fig.\ref{fig4} (Left), pertaining to the 30-40\% collision centrality, we observe the relationship between the 3-particle  $\gamma$ correlator and $f_{DbCS}$ for an injected CME signal of approximately 1\%, as well as for AMPT and charge-shuffled (ChS) samples, considering opposite-sign (OS) and same-sign (SS) charge pairs. The $\gamma$ magnitude significantly increases for both SS and OS charge pairs in the top $f_{DbCS}$ bins compared to the average value for a given collision centrality, as depicted in Fig.\ref{fig2} (Top Right). In particular, opposite-sign charge pairs exhibit weak positive correlations ($\gamma$ $>$ 0), while same-sign charge pairs display strong negative correlations ($\gamma$ $<$ 0) in the top $f_{DbCS}$ bins, indicating a back-to-back charge separation phenomenon in these events. As $f_{DbCS}$ values decrease, correlations diminish and approach zero for lower $f_{DbCS}$ bins. Notably, $\gamma_{SS}$ is negative while $\gamma_{OS}$ is positive for the top 0-40\% $f_{DbCS}$ bins, reflecting the back-to-back charge separation. Similar trends are observed across different centralities, with fewer (more) top $f_{DbCS}$ bins for higher (lower) collision centralities. In Fig.\ref{fig4} (Right), the $\delta$ correlator plots for various $f_{DbCS}$ bins for the same collision centrality are displayed for opposite- and same-sign charge pairs. Here again, the $\delta$ values are significantly enhanced for the top $f_{DbCS}$ bins compared to the overall values (Fig.\ref{fig2} (Top Left)), with enhancement levels similar to those observed for $\gamma$. Additionally, Fig.\ref{fig4} illustrates the independence of correlated backgrounds (i.e., $Corr_{AMPT}$ and $Corr_{CME}$) on $f_{DbCS}$. It is noted that correlated backgrounds remain consistent across different $f_{DbCS}$ bins, a trend observed consistently across various collision centralities.\par
\begin{figure*}[h!]
  \centering{
  \includegraphics[width=.85\columnwidth]{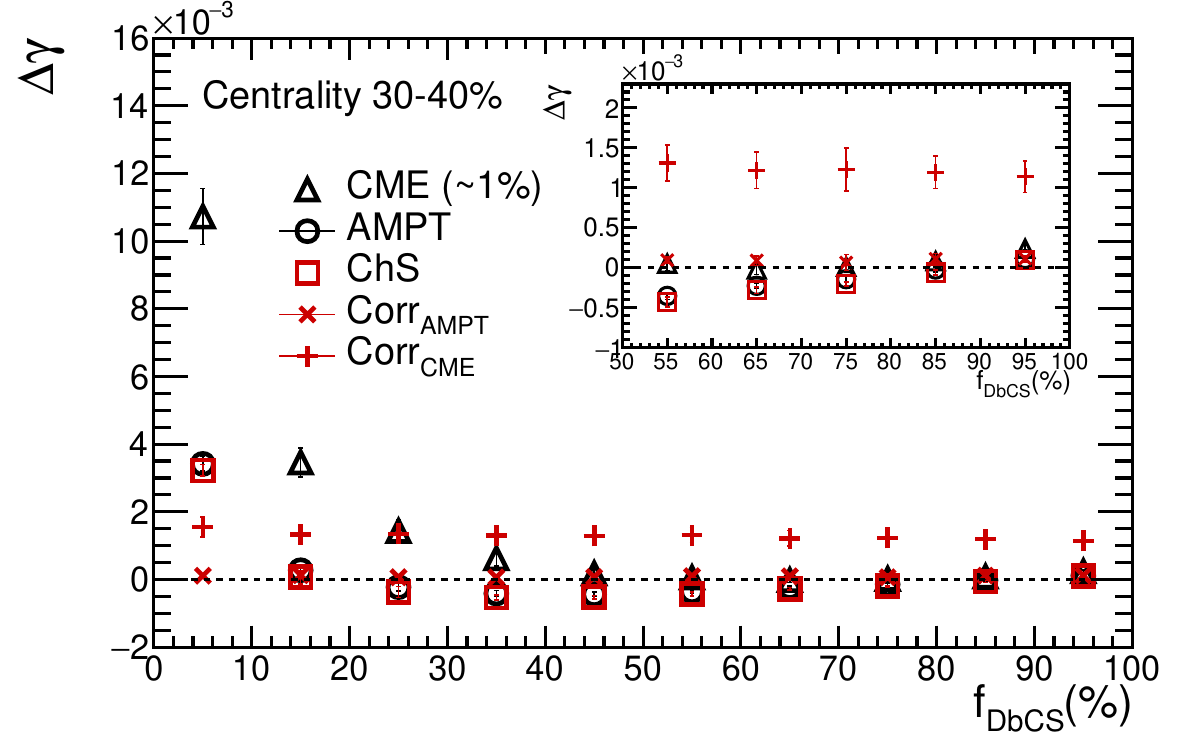}
  \includegraphics[width=.85\columnwidth]{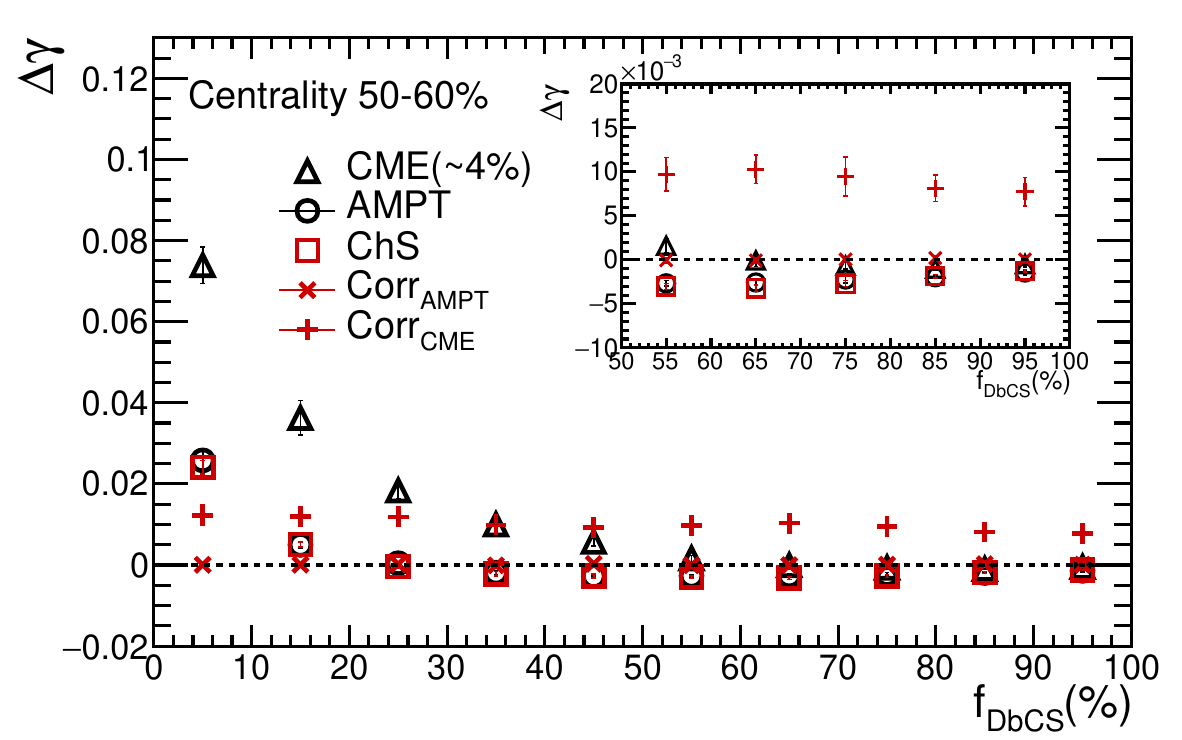}
  \includegraphics[width=.85\columnwidth]{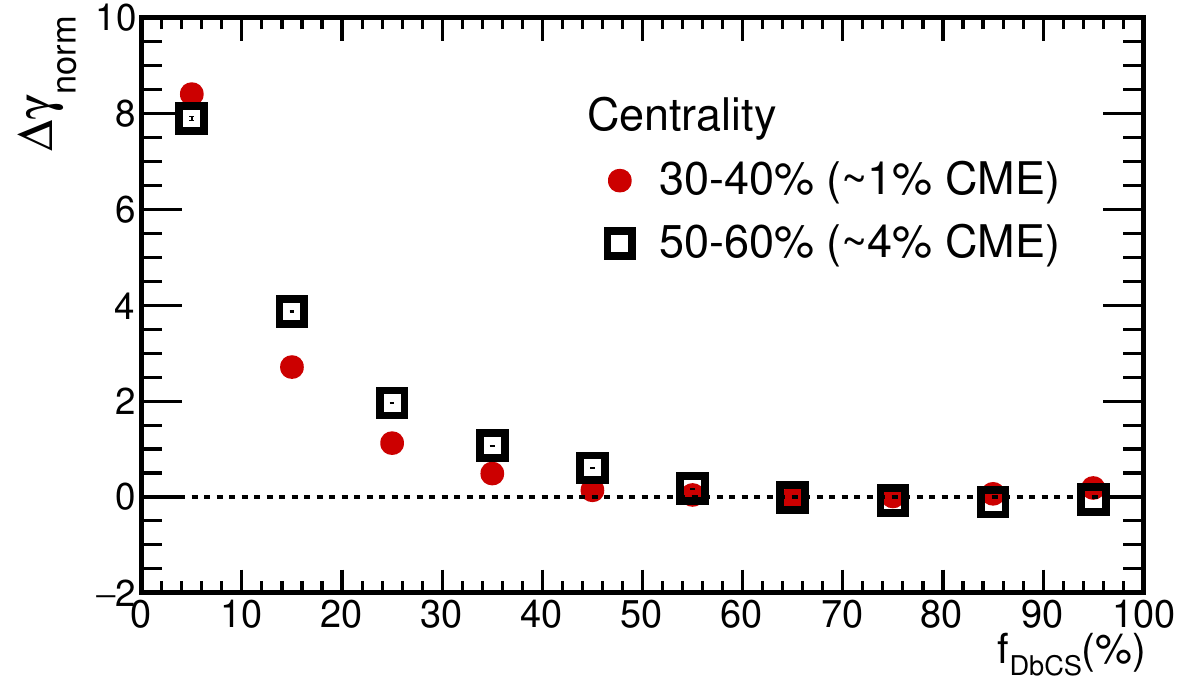}}
  \caption{Top Left: The dependence of $\Delta\gamma$ on $f_{DbCS}$ for the 30-40\% collision centrality for AMPT, CME ($\sim$1\% CME) and charge-shuffled (ChS) alongwith corresponding correlated backgrounds for the CME ($Corr_{CME}$) and AMPT ($Corr_{AMPT}$) as indicated in the figure. Inset shows the magnified $\Delta\gamma$ for 50-100\% $f_{DbCS}$. Top Right: Similar to Top Left plot but for the 50-60\% ($\sim$ 4\% CME) collision centrality. Bottom: The dependence of $\Delta\gamma_{norm}$ on $f_{DbCS}$ for 30-40\% and 50-60\% collision centralities.}
  \label{fig7}
\end{figure*}

\begin{table*}[!htb]
  \begin{center}
\resizebox{\textwidth}{!}{%
\begin{tabular}{|c|c|c|c|c|}
    \hline
    \multirow{2}{*}{\large{$f_{DbCS}$}} & \multicolumn{2}{c|}{\large CME} & \multicolumn{2}{c|}{\large AMPT} \\[2pt]
    \cline{2-5}
    & {50-60\% Centrality} & {30-40\% Centrality} & {50-60\% Centrality} & {30-40\% Centrality} \\
    & ($\sim$4\% CME) & ($\sim$1\% CME) & (No CME) & (No CME) \\ [1.5pt]
    \hline
    0-10\% & 0.508$\pm$0.079 & 0.555$\pm$0.094 & 0.01$\pm$0.124 & 0.045$\pm$0.081 \\[1pt]
    \hline
    10-20\% & 0.533$\pm$0.146 & 0.615$\pm$0.175 & 0.07$\pm$0.238 & - \\[1pt]
    \hline
    20-30\% & 0.354$\pm$0.155 & 0.061$\pm$0.304 & - & - \\[1pt]
    \hline
    30-40\% & 0.010$\pm$0.246 & - & - & - \\
    \hline
\end{tabular}}
\end{center}
  \caption{The $f_{CME}$ values for different $f_{DbCS}$ bins with $\Delta\gamma_{norm} > 1$ for 50-60\% and 30-40\% centralities for AMPT and CME samples.}
  \label{table1}
\end{table*}

We have also computed in-plane and out-of-plane correlations for both SS and OS charge pairs for different $f_{DbCS}$ bins to confirm whether or not the observed correlations are out-of-plane as expected in case of the CME. Fig.\ref{fig5} displays the dependence of in- and out-of-planes correlations on the charge separation ($f_{DbCS}$) for different charge combinations for the 30-40\% collision centrality. Correlations for the top 0-40\% $f_{DbCS}$ are out-of-plane whereas in-plane weak correlations are observed as expected for the CME-like correlations due to out-of-plane charge separation. These correlations decrease with decreasing $f_{DbCS}$ values. Similar trends are also observed for other collision centralities resulting increase/decrease in correlation magnitude with decreasing/increasing collision centrality.\par
It is noted that in the CME-enriched samples, $\sqrt{|\gamma_{SS}|}$ for the top 10\% $f_{DbCS}$ events from each collision centrality exhibits a linear variation with positive charge asymmetry ($\langle A^{+} \rangle$) across the dumbbell (Fig.\ref{fig6}), as anticipated (as per Eq.(\ref{eq:eq2})), given that ($\langle A^{+} \rangle$) is directly proportional to the parity-violating parameter $a_{1}$ (i.e., A$^+$ = $\pi a_{1}/4$)~\cite{Kha05}. The data points representing $\sqrt{| \gamma_{SS} |}$ versus $\langle A^+ \rangle$ are fitted with a straight line, $\sqrt{|\gamma_{SS}|}=p_{0}+p_{1} \langle A^{+} \rangle $. For the CME samples, the fitted values of $p_0$ and $p_1$ are -0.077 $\pm$ 0.00702 and 1.04 $\pm$ 0.0502, respectively, with a total $\chi^2$ value is 1.22.\par
Fig.\ref{fig7} (Top Left) and Fig.\ref{fig7} (Top Right) present the $\Delta\gamma$ versus charge separation ($f_{DbCS}$) plots for 30-40\% ($\sim$1\% CME) and 50-60\% ($\sim$4\% CME) collision centralities, respectively. The values of $\Delta\gamma$ for correlated background (Corr) and charge-shuffled (ChS) are also displayed in these figures. It is observed that charge-shuffled and AMPT data points agree within statistical uncertainties for various $f_{DbCS}$ bins whereas the data points corresponding to the CME samples have larger correlation values. It is also seen that the $\Delta\gamma$ for the CME samples have larger values than those of combined charge-shuffled and correlated backgrounds for the top $f_{DbCS}$ bins. The variation of $\Delta\gamma$ normalized to the average value ($\Delta\gamma_{avg}$) for a given collision centrality (i.e., $\Delta\gamma_{norm}$ = $\Delta\gamma/\Delta\gamma_{avg}$) versus charge separation ($f_{DbCS}$) is displayed in Fig.\ref{fig7} (bottom) for 50-60\% and 30-40\% collision centralities. It is seen that $\Delta\gamma_{norm}$ is greater than 1 for 50-60\% centrality (corresponding to $\sim$4\% injected CME signal) for the top 40\% $f_{DbCS}$ whereas it is greater than 1 for the top 30\% $f_{DbCS}$ for 30-40\% centrality (i.e., injected CME signal $\sim$1\%). Further, it is seen that for the top 10\% $f_{DbCS}$ the $\Delta\gamma_{norm}$ is $\sim$8 and it decreases rapidly with decreasing externally injected CME signal and $\Delta\gamma_{norm}$ approaches zero for $f_{DbCS}$ 90-100\%. For AMPT and charge-shuffled samples similar type of enhancement in $\Delta\gamma$ can be seen in Fig.\ref{fig7} (top left and top right), but not shown in Fig.\ref{fig7} (bottom) for the sake of clarity as their average values are too small (Fig.\ref{fig2} Bottom). \par
An attempt is made to get the fractional CME ($f_{CME}$) contribution in the observed $\Delta\gamma$ for the top $f_{DbCS}$ bins wherein the $\Delta\gamma_{norm}>1$ using the following equation:
\begin{equation}
  f_{CME} = \frac{(\Delta\gamma - \Delta\gamma_{bkg})} {\Delta\gamma}
  \label{eq:eq12}
\end{equation}
Table~\ref{table1} presents the fraction of Chiral Magnetic Effect (CME) contributions for the top $f_{DbCS}$ bins in 50-60\% collision centrality (with approximately 4\% injected CME signal) and 30-40\% collision centrality (with approximately 1\% injected CME signal). These bins exhibit $\Delta\gamma_{norm} > 1$ for both AMPT and CME samples. Notably, the fraction is significant in the top $f_{DbCS}$ bins, and the frequency of bins showing this trend decreases as the percentage of externally injected CME-like signal decreases. However, the extracted fraction of CME ($f_{CME}$) is approximately zero within statistical errors for the AMPT samples lacking externally injected CME-like signal. These findings underscore the importance of focusing on the top $f_{DbCS}$ bins with the highest back-to-back charge separation for a given collision centrality to confirm the presence of CME in heavy-ion collisions.
\section{Summary}
We conducted an analysis of Au+Au events generated by AMPT with string melting configuration, at a center-of-mass energy $\sqrt{s}_{\mathrm{NN}}=200$ GeV. Additionally, we examined AMPT samples with externally injected Chiral Magnetic Effect (CME) signal using the sliding dumbbell method. For the first time, we successfully isolated events that exhibit CME-like signal, corresponding to the top 10\% of dumbbell charge separation ($f_{DbCS}$) bins. In these events, the CME-sensitive 3-particle $\gamma$ correlator showed a significant enhancement, as demonstrated in Fig.\ref{fig4}, resulting in an approximately 8-fold increase in the value of $\Delta\gamma$ for the top $f_{DbCS}$ bins (Fig.\ref{fig7} Bottom).\par
We observed that the $\sqrt{| \gamma_{SS} |}$ value exhibits a linear relationship with positive charge asymmetry ($\langle A^+ \rangle$) across the dumbbell. Consequently, we were able to access the parity-violating parameter, $a_1$, through $\langle A^+ \rangle$ for the first time. The fraction of CME extracted from the top $f_{DbCS}$ bins was found to be substantial ($>$0.5), while the extracted CME fraction from the AMPT sample was statistically zero. Therefore, the sliding dumbbell method enables the detection of even the smallest CME signal (approximately 1\%).\par
Given these results, it is feasible to employ the Sliding Dumbbell Method in future studies of Au+Au, isobaric (Zr+Zr, Ru+Ru) collisions at RHIC, and Pb+Pb collisions at LHC energies to meticulously search for the CME signal.
\section*{Acknowledgments}
The financial assistance from the Department of Science $\&$ Technology and University Grants Commission, Government of India, are gratefully acknowledged. The authors are also thankful to the Panjab University and the Physics Department for providing the research facilities.

\end{document}